\documentclass[reprint,amssymb,amsmath,aip]{revtex4-1}
\usepackage{amsmath}
\usepackage{graphicx}

\begin{document}

\title{The impact of hysteresis on the electrocaloric effect at first-order 
phase transitions}

\author{Madhura Marathe}
\author{Claude Ederer}
\affiliation{Materials Theory, ETH Z\"urich, Wolfgang-Pauli-Str. 27, 8093 Z\"urich, Switzerland}
\author{Anna Gr\"unebohm}
\email{anna@thp.uni-due.de}
\affiliation{Faculty of Physics and CENIDE, University of Duisburg-Essen, 47048 Duisburg, Germany}


\begin{abstract} We study the impact of thermal hysteresis at the first-order
  structural/ferroelectric phase transitions on the electrocaloric
  response in bulk BaTiO$_3$ by performing molecular dynamics
  simulations for a first-principles-based effective Hamiltonian. We
  demonstrate that the electrocaloric response can conceptually be
  separated in two contributions: a \emph{transitional} part, stemming
  from the discontinuous jump in entropy at the first order phase
  transition, and a \emph{configurational} part, due to the continuous
  change of polarization and entropy within each phase. This latter
  part increases with the strength of the applied field, but for small
  fields it is very small. In contrast, we find a large temperature
  change of $\sim 1$\,K resulting from the transition entropy, which
  is essentially independent of the field strength.
  However, due to the coexistence region close to the first order
  phase transition, this large electrocaloric response depends on the
  thermal history of the sample and is generally not reversible. We
  show that this irreversibility can be overcome by using larger
  fields.
  \end{abstract}
\maketitle   

\section{Introduction}
The electrocaloric (EC) effect results in an adiabatic temperature
change (or an isothermal entropy change) with variation of an external
electric field.\cite{Scott_2011,Valant_2012}  Along with the
magnetocaloric and elastocaloric effects, the EC effect has the
potential to lead to energy-efficient and environmentally friendly
solid-state cooling
devices.\cite{Moya/Narayan/Mathur:2014,Faehler_et_al:2011} 
Typically, the largest caloric response is observed near ferroic phase
transitions.\cite{Moya/Narayan/Mathur:2014} In particular, giant
temperature changes have been found at first-order (FO) phase
transitions with coupled ferroic and structural degrees of
freedom.\cite{Liu2,Titov} However, close to FO transitions, a
coexistence region exists, which leads to thermal hysteresis, i.e.,
within a certain temperature range, the system can exist in different
(meta-)stable states depending on the history of the sample. Thus, at
the same temperature and applied field strength, the state of the
system can differ in field cooled (FC), field heated (FH), zero-field
cooled (ZFC), or zero-field heated (ZFH) samples.

The impact of thermal hysteresis on the reversibility of the caloric
response has been studied extensively for the case of magnetic materials
close to magneto-structural phase
transitions.\cite{Faehler_et_al:2011,Liu2,Titov,Gutfleisch} It has
been shown that for systems with broad hysteresis, a giant caloric
response can typically only be found for the first field
pulse. However, for cooling applications, a device needs to be
operated over a large number of field cycles. Therefore, the search
for hysteresis-free materials, ways to bypass the thermal hysteresis,
or the use of smaller reversible responses within the coexistence
region is a very active area in the field of magnetic 
materials.\cite{Liu2,Titov,Gutfleisch}  In contrast, there are only
few such studies for the EC effect. Thus, a better understanding of
the impact of thermal hysteresis is needed for future applications
of EC materials.

The EC effect has been extensively studied in BaTiO$_3$ (BTO), see,
e.g.,~\onlinecite{Narayan/Mathur:2010,Moya_et_al_2013,Novak_Pirc_Kutnjak:2013,Akcay_et_al_2007,Novak_Kutnjak_Pirc:2013,Beckman_et_al_2012,Nishimatsu_Barr_Beckman:2013,Marathe/Ederer:2014,Grunebohm_Nishimatsu:2016,Marathe_et_al:2016}. BTO
exhibits a paraelectric (PE) cubic (C) phase at high temperatures and,
on cooling, undergoes three transitions to ferroelectric (FE)
tetragonal (T), orthorhombic (O), and rhombohedral (R) phases. In
absence of an external field, each FE transition in BTO is a FO
transition.\cite{Lines-Glass} For small fields, this results in
thermal hysteresis and coexistence regions, in which the state of the
system depends on the preceding heat treatment. Previous studies have
shown that under increasing strength of an applied field, the
transition temperatures are shifted, the thermal hysteresis is
reduced, and for sufficiently large fields the nature of the
transition can
change.\cite{Bell:2001,Li_Cross_Chen:2005,Marathe_et_al:2017} For the
PE-FE transition, there is no well-defined phase transition any more
above a critical field strength $E_{\text{c}}$.\cite{Novak_Pirc_Kutnjak:2013}

Some experimental measurements performed at small applied fields have
indeed reported irreversibility in the EC response of BTO. For
example, Moya, \textit{et al.}\cite{Moya_et_al_2013} have found an
irreversible EC response at the PE-FE transition. Furthermore, around
the FE-FE (T-O) transition, Bai, \textit{et al.}\cite{Bai_et_al:2012}
measured a large EC response only during the first application of an
electric field, which was then reduced to a rather small response on
further cycling, analogous to the irreversibility
found at magneto-structural phase transitions.\cite{Titov,Gutfleisch}

Most theoretical studies so far have focused on the EC effect close to
the PE-FE transition under large fields. Since at large fields ($E >
E_{\text{c}}$) there is no FO phase
transition,\cite{Novak_Pirc_Kutnjak:2013,Marathe_et_al:2017} 
irreversible effects are not expected in this case. On the other hand,
an irreversible EC effect has been predicted even for large fields,
when different metastable states occur in defect-doped
BTO.\cite{Grunebohm_Nishimatsu:2016} Under small fields, thermal
hysteresis exists at all transitions, and thus an irreversible EC
effect is expected, which depends strongly on the thermal history of
the sample.

In this paper, we study the EC response of BTO at applied fields that
are smaller than the critical fields. In particular, we examine how
the EC effect depends on the thermal history of the sample, and
whether periodic cycling of the field near the coexistence region
results in irreversibilities.
To this end, we perform molecular dynamics (MD) simulations for an
effective Hamiltonian derived from first principles (see, e.g.,
Ref.~\onlinecite{Marathe_et_al:2016}). We focus on the EC response
around the C-T and T-O transitions. We first show that, within the
coexistence region, the EC temperature change indeed depends on the
initial phase of the system. Then, we study the EC response for
different applied field strengths around the critical field while
switching the field on and off. Further, we apply a field for up to
two cycles at a few selected temperatures to examine the reversibility
of the response. We observe that the effect is irreversible for small
fields, with a large response in the first cycle, which then reduces
by an order of magnitude in the following cycle. Finally, we  briefly compare
the observed behavior to the well-known hysteretic effects in materials showing a FO magneto-structural phase transition.

\section{Computational details}
\label{sec:comp-details}

\begin{figure}[tb]
\centering \includegraphics*[width=.5\textwidth]{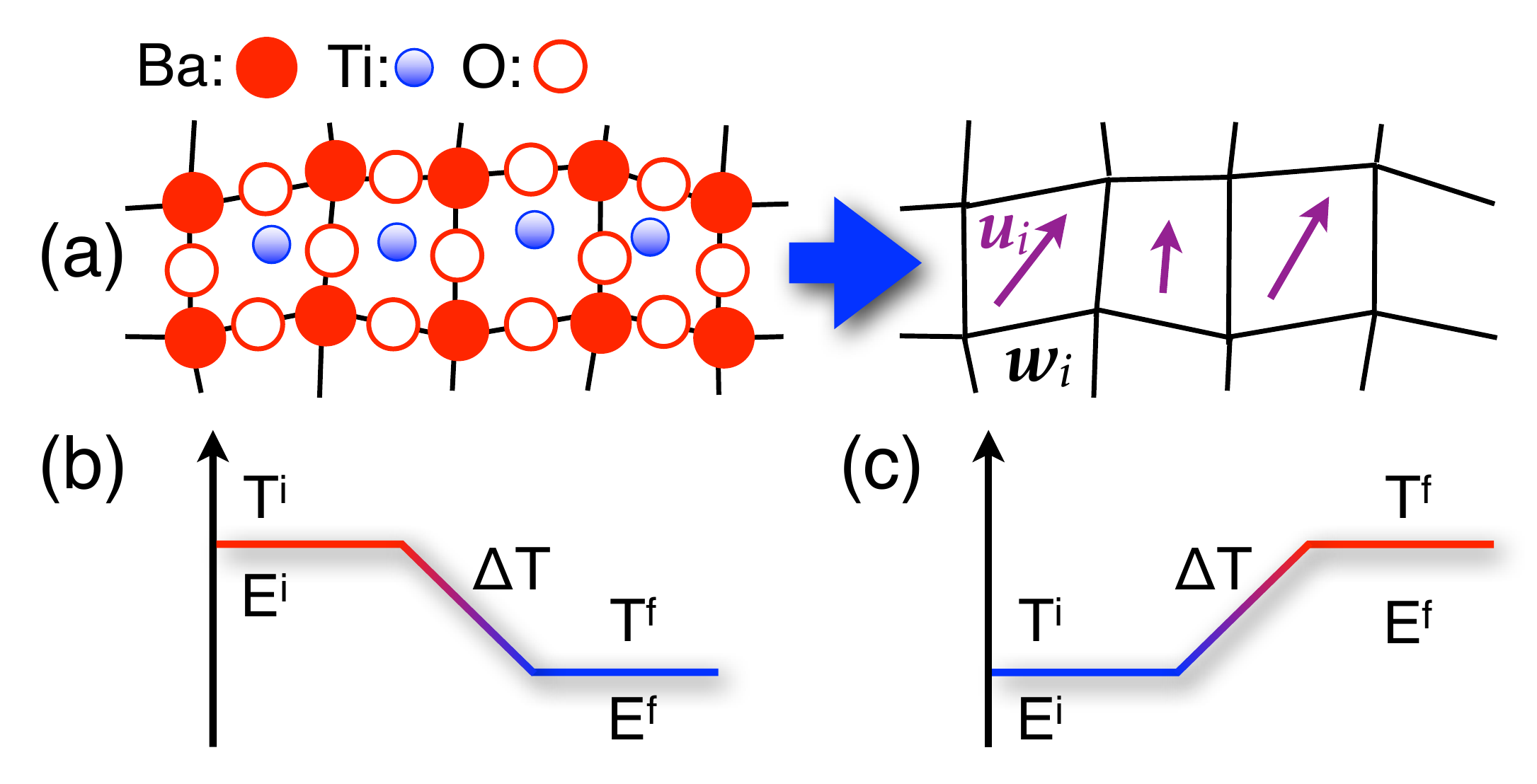}
  \caption[]{(a) Illustration of the construction of the effective
    Hamiltonian: the local atomic arrangement is mapped on the local
    soft mode vectors $\bf{u}_i$ and the local strains $\bf{w}_i$ in
    each unit cell. (b)-(c) Illustration of the direct simulation of
    the EC temperature change: the system is equilibrated in an
    external field $E_i$ at the start temperature $T_i$, the field is
    ramped up or down to the final field $E_f$ and after equilibrium
    is reached, the final temperature $T_f$ is recorded. $\Delta T$
    indicates the adiabatic temperature change on varying the
    field. For normal EC response, $\Delta T$ is negative when the
    field is ramped down (b) and positive when the field is ramped up
    (c).
  }
    \label{fig:ramping}
\end{figure}

To construct the effective Hamiltonian used in our work, the atomic
degrees of freedom in each unit cell $i$ are mapped on a ferroelectric
soft mode vector (${\bf u}_i$) and a local strain (${\bf w}_i$), see
Fig.~\ref{fig:ramping}~(a).\cite{Zhong_Vanderbilt_Rabe_1994,Zhong_Vanderbilt_Rabe_1995} The
total energy of the system is then expressed as a low order polynomial
in terms of these variables. All parameters of this effective
Hamiltonian have been calculated using \textit{ab initio} density
functional theory.\cite{Nishimatsu_et_al_2010} We use the open-source
{\it{feram}} code
(http://loto. sourceforge.net/feram/)\cite{Nishimatsu_et_al_2008},
which allows to perform MD simulations for the effective Hamiltonian
at finite temperatures and fields. We use a $96 \times 96 \times 96$
supercell to simulate bulk BTO within periodic boundary
conditions. For computational efficiency, we treat only the soft mode
variables as dynamical variables whereas the local and global strains
are optimized in each MD step according to the current soft mode
configuration.

This method has been successfully used to calculate the EC response,
see
e.g.,~\onlinecite{Beckman_et_al_2012,Nishimatsu_Barr_Beckman:2013,Marathe/Ederer:2014,Grunebohm_Nishimatsu:2016,Marathe_et_al:2016,Ponomareva/Lisenkov:2012}. We
calculate the response in bulk BTO using the direct method.
First, we allow the system to equilibrate at a given initial
temperature $T_\text{i}$ and initial field $E_{\text{i}}$ using
sufficiently long thermalization times in the MD simulations.
Then, the applied electric field is either switched off (ramp down) or
on (ramp up), as illustrated in Figs.~\ref{fig:ramping}(b) and (c)
respectively. In this study, we always apply the field along the
pseudo-cubic [001] direction.
During and after the field-ramping, we allow the system to evolve and
then equilibrate again within the constant energy (microcanonical)
ensemble before averaging the physical quantities of interest. The
corresponding EC temperature change $\Delta T$ is obtained as
difference between the initial and final temperatures.
Note that the effective Hamiltonian results in a reduction of degrees
of freedom from 15 in the real system to 3 in the model system.
Therefore, the change in temperature is rescaled by a factor of 1/5.
More details of our calculations are described in
Refs.~\onlinecite{Marathe_et_al:2016,Marathe_et_al:2017}.

To study the impact of thermal hysteresis, i.e., the thermal history
of the sample, we prepare the system by either heating or cooling
simulations under a constant external field. At each temperature $T$,
local dipole configurations are obtained after thermalization and then
used as initial configuration to calculate the EC temperature change.
In addition, we have also performed simulations where the starting
configuration was obtained by simply randomly initializing the local
dipoles and subsequent thermalization at a given temperature and
field, ensuring that the system is in a well-defined initial state,
before ramping the field either up or down from a given initial
temperature and monitoring the asymmetry of the obtained EC
temperature change.

\begin{figure*}[tb]
\centering
\includegraphics[width=0.82\textwidth]{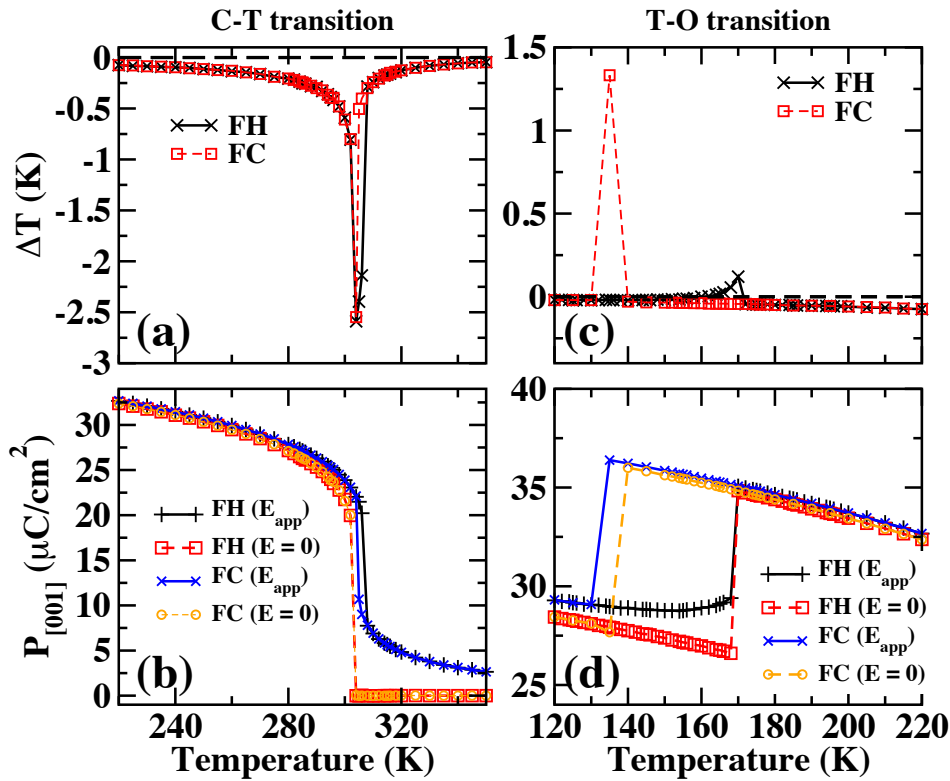}
\caption{The EC temperature change is plotted as a function of initial
  temperature for an applied field of 20\,kV/cm around the C-T
  transition (a) and the T-O transition (c). Two differently prepared
  initial configurations (FC and FH) are used at each temperature, 
  leading to different EC response for the same applied field. The
  corresponding polarization components along the applied field
  direction, $P_{[001]}$, are plotted in (b) and (d) before (solid
  lines) and after (dashed lines) the field has been removed.}
  \label{fig:delT_T_E20}
\end{figure*}

\section{Results and discussion}
First, we investigate the influence of the thermal history on the EC
effect of BTO. We calculate the EC response for ramping down a field
of $E_\text{i} =$ 20\,kV/cm applied along [001] near the C-T and T-O
transitions. This field strength is below the critical field strength
found for our model and thus there is thermal hysteresis at both
transitions.\cite{Marathe_et_al:2017} To determine the influence of
the thermal history, we perform two simulations for each $T_\text{i}$
using the two starting configurations obtained from FH and FC
simulations.

The calculated $\Delta T$ as function of initial temperature is
plotted in Fig.~\ref{fig:delT_T_E20}(a) and (c). We observe a clear
difference between the EC response calculated for the FH and FC
starting configurations. At the PE-FE (C-T) transition, we obtain a
peak value of $|\Delta T| \sim 2.5$\,K at 304\,K for both FC and FH
starting configurations. However, the peak is sharper (only one data
point) for the FC configurations as compared to the FH case, for which
the peak extends up to $T_\text{i}=306$\,K.
At the T-O transition, we obtain a sharp peak at 135\,K
for the FC case with $|\Delta T| \sim 1.3$\,K. The response is
inverse, i.e., positive $\Delta T$ under field removal, as expected
for this transition and field along [001], see
Ref.~\onlinecite{Marathe_et_al:2017}. For the FH case we observe only a
small feature around 170\,K with a magnitude of $\sim $0.12\,K and a
broad decrease towards the low-T side.

To better understand these differences between the FC and FH response,
Figs.~\ref{fig:delT_T_E20}(b) and (d) show the components of the
electric polarization along the applied field direction as function
of initial temperature for the initial and final states,
respectively, i.e., with and without applied field. One recognizes
that the difference in the FC and FH EC response is related to
differences in the initial and/or final states of the system.
At the PE-FE transition, the FH initial configurations are in the FE-T
phase up to $T_\text{i} = 306$\,K, whereas in the FC case, the system
is initially in the FE-T phase only for $T_\text{i} \le$
304\,K. Consequently, the corresponding EC response differs for
$304\,\text{K} < T_\text{i} \le 306\,\text{K}$.
The final states after field removal are the same in both cases, for
$T_\text{i} < 304$\,K the system ends up in the FE phase, whereas for
$T_\text{i} \ge 304$\,K it ends up in the PE phase with zero
polarization.
Thus, one also recognizes that the large EC response with $|\Delta T|
> 2$\,K occurs only in those cases where the system undergoes a
transition from the FE to the PE phase under field removal. In all
other cases the removal of the electric field only leads to a
reduction or vanishing of (induced) polarization, but the system does
not undergo a phase transition. In these cases the EC response is much
smaller ($|\Delta T| \le 1$\,K).

Conceptually, the EC response can therefore be divided into two
different contributions -- (i) a \emph{configurational} part,
resulting from continuous changes of polarization and entropy induced
by the applied field without triggering a phase transition,
and (ii) a \emph{transitional} part related to the discontinuous jump
of entropy related to the field-induced FO phase transition. The
analysis of Fig.~\ref{fig:delT_T_E20}(a) and (b) shows that for fields
of the order of 20\,kV/cm the large EC response of BTO can mainly be
attributed to the transitional part (ii), whereas the configurational
part (i) is very small.

The same holds true at the T-O transition. For the FC simulations, the
large peak in $\Delta T$ occurs at $T_\text{i} = 134$\,K, i.e., on the
low temperature side of the coexistence region, when the system can
undergo a phase transition from T to O under field removal. This can
be seen from the polarization components in the initial and final
states shown in Fig.~\ref{fig:delT_T_E20}(d). At all other
temperatures, no phase transition occurs, and the EC response is
negligibly small.
For the FH starting configurations, the system is initially in the O
phase up to the (field-dependent) high-T side of the
coexistence region. Since an applied field of 20 kV/cm is not sufficient
to shift this phase boundary below the ZFC transition
temperature,\cite{Marathe_et_al:2017} the system remains in the O
phase under field removal. Thus, no phase transition occurs and the EC
response remains small over the whole temperature region around the
T-O transition.  The small feature in $\Delta T$ below 170\,K, i.e.,
just below the FH T-O transition, is due to an enhanced polarization induced by the applied field.
We note that the field-dependent phase transition lines are rather steep (see
Ref.~\onlinecite{Marathe_et_al:2017}) and thus the temperature ranges where
a field-induced transition can occur under small fields are extremely
narrow. Thus, whether we obtain one or more data points with a phase
transition strongly depends on the $T$-sampling.

\begin{figure}[tb]
\centering
\includegraphics*[width=0.45\textwidth]{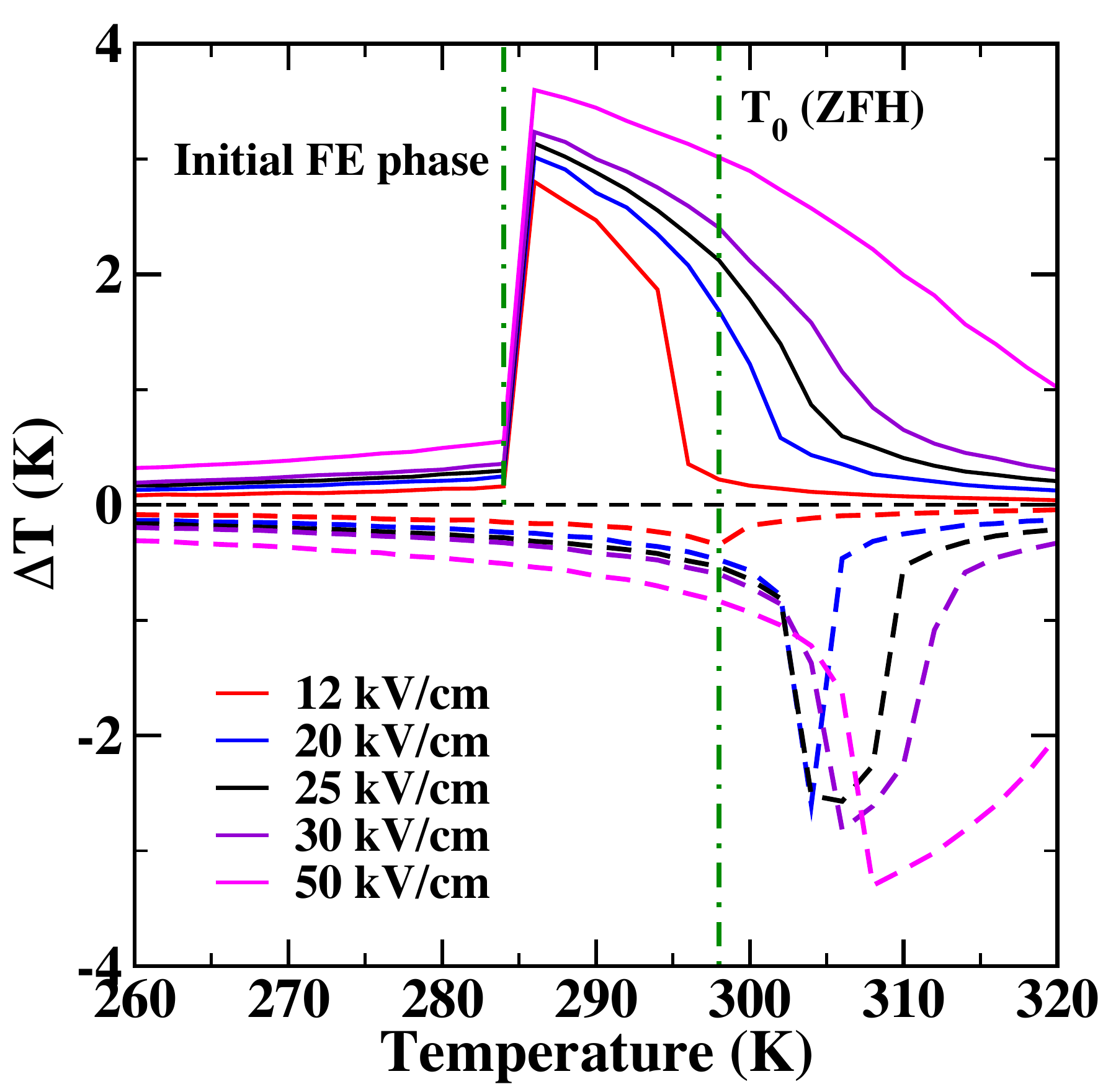}
\caption{EC temperature change as function of the initial temperature $T_\text{i}$
  corresponding to application (solid lines) and removal (dashed
  lines) of the field for different applied field strengths. In these
  simulations, the on/off cases are calculated independently based
  on randomly initialized starting configurations. The ZFH transition
  temperature $T_0(\text{ZFH})$ is indicated by the vertical
  dot-dashed line. The other vertical dot-dashed line at $T=284$\,K
  indicates the temperature where the initial state for the ``on''
  case changes from FE at lower temperatures to  PE at higher temperatures.}
\label{fig:ece}
\end{figure}

\begin{figure}[tb]
\centering
\includegraphics*[width=0.49\textwidth]{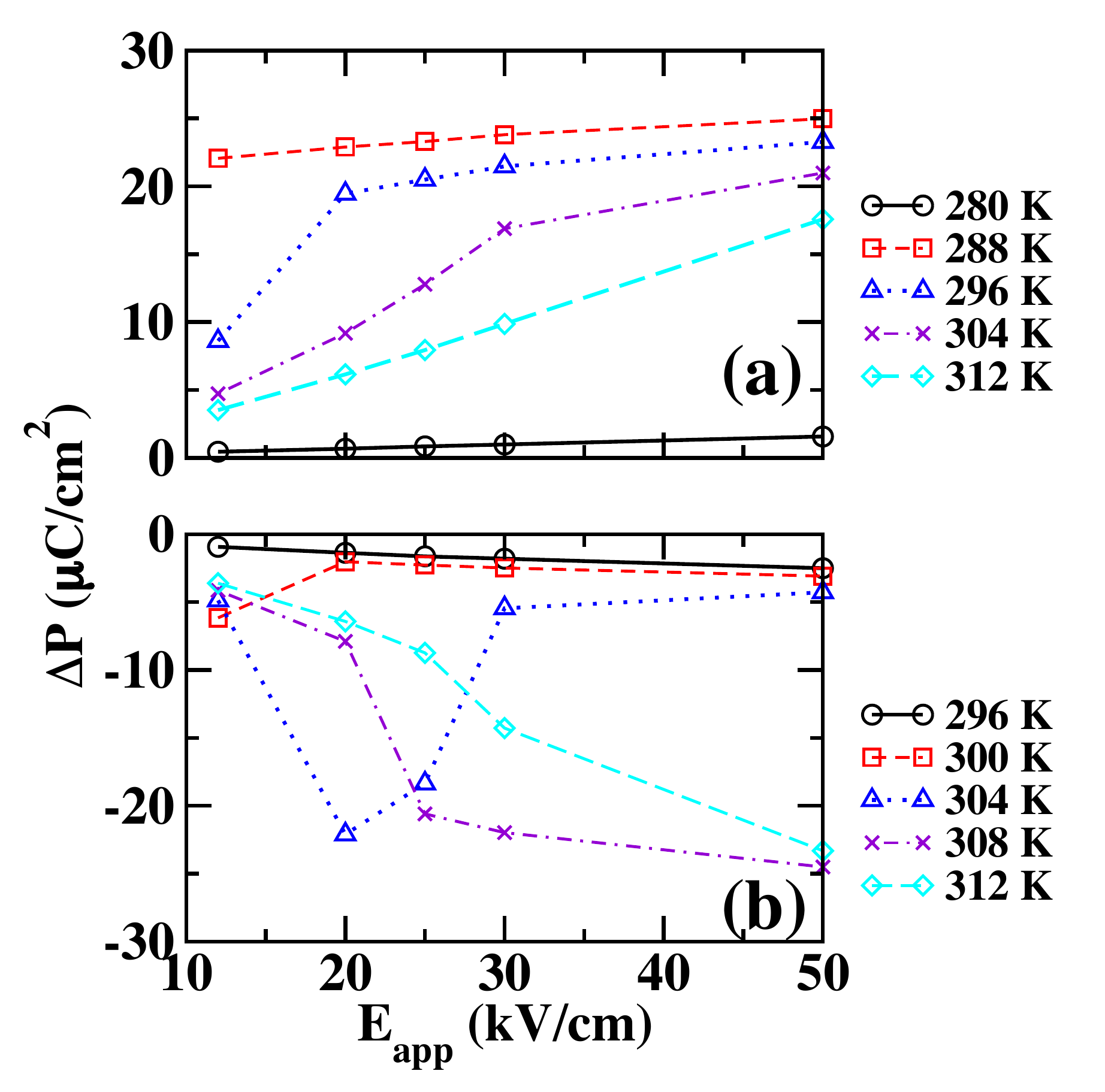}
\caption{Change in polarization, $\Delta P$, between final and initial
  states for the on (a) and off (b) ramping of the field for different applied field strengths
    at a few selected temperatures corresponding to the caloric response in Fig.~\ref{fig:ece}.}
\label{fig:pol}
\end{figure}

Up to now, we have examined how the thermal hysteresis affects the EC
response at both C-T and T-O transitions. Next, we focus on the PE-FE
(C-T) transition and compare the EC effect under field application and
removal, respectively, thereby also considering the impact of the
strength of the applied field. In this case, we use starting
configurations which have been obtained after random initialization of
the local dipoles and subsequent thermalization at the initial
temperature $T_\text{i}$, either with or without applied field.
Fig.~\ref{fig:ece} shows the resulting EC temperature change as
function of $T_\text{i}$ for several applied field strengths. The
data-sets with positive $\Delta T$ correspond to application of the
field (``on''), whereas the data-sets with negative $\Delta T$
correspond to removal of the field (``off''). Two main features can be
recognized: i) for small fields ($E_\text{app} < 50$\,kV/cm) the
magnitude and shape of the peaks in the EC effect differ significantly
between the on and off cases, and ii) for larger fields ($E_\text{app}
= 50$\,kV/cm) the $\Delta T$ curves look similar for the on/off cases
but are shifted against each other on the temperature axis. This
latter effect can easily be understood from the condition for a
completely reversible EC cycle (see Ref.~\onlinecite{Marathe_et_al:2016}):
\begin{equation}\label{eq:EC-cond}
\Delta T_{\text{off}}(T_\text{i} + \Delta T_{\text{on}}(T_\text{i})) = -\Delta T_{\text{on}}(T_\text{i}).
\end{equation}
Here, $\Delta T_{\text{off}}$ and $\Delta T_{\text{on}}$ correspond to
the EC temperature changes under field removal and application,
respectively. Note, however, that Eq.~\eqref{eq:EC-cond} is not
directly applicable to the data shown in Fig.~\ref{fig:ece}(a), since
$\Delta T$ has been corrected according to the reduced number of
degrees of freedom within the effective Hamiltonian (see
Sec.~\ref{sec:comp-details}). Therefore, the shift in the peak
positions in Fig.~\ref{fig:ece}(a) appears to be much larger than the
magnitude of $\Delta T$.

To better understand the difference between the on and off cases for
small applied fields, we also analyze the field-induced change in
polarization, $\Delta P$:
\begin{equation}
\Delta P = P_f (T_f,E_f)- P_i(T_i,E_i) \quad .
\end{equation}
Here, $P_i$ and $P_f$ denote the polarization in the initial and final
states, i.e., before and after the field
ramping. Fig.~\ref{fig:pol}(a) and (b) depict $\Delta P$ as function
of applied field for a few selected temperatures around the peak in
$\Delta T$ for the on and off cases, respectively.
We first note that for the on case, our initialization (with
$E_\text{i}=0$) has put the system in the FE phase for all
temperatures below $T_\text{i}=284$\,K and in the PE state for all
temperatures above that. Thus, below $284$\,K, application of an
electric field leads only to minute changes in polarization (see,
e.g., data for $T=280$\,K in Fig.~\ref{fig:pol}(a)) and consequently
the EC effect is very small at this temperature. At slightly higher temperatures, e.g.,
$T=288$\,K, even a small field of 12\,kV/cm can trigger the FO phase
transition to the FE state, leading to a steep rise in $\Delta T$ due
to the large transitional contribution to the EC effect.~\footnote{We
  note that for a ZFC starting configuration, the steep rise on the
  low-$T$ side of the EC peak occurs exactly at the ZFC transition
  temperature.}  Further increasing the field strength at this
temperature results only in a small further increase of $\Delta T$,
due to an increased configurational contribution.
At temperatures close to the ZFH transition temperature, e.g.,
$T = 296$\,K, a small field of 12\,kV/cm is not sufficient to trigger
the transition to the FE phase (note that here the system is within
the coexistence region), and thus the EC peak is cut off on the
high-$T$ side. For higher temperatures, above the coexistence region,
the EC effect increases monotonously with the applied field
strength. This is due to the high polarizability of the PE phase in
this temperature and field region close to the phase transition, which
leads to an increasingly large contribution from the configurational
part and a broadening of the EC response towards higher temperatures.
The characteristic peak shape with a sharp drop towards the low-T side
and a field-dependent broad shoulder towards higher temperatures has
also been observed experimentally in
Ref.~\onlinecite{Novak_Kutnjak_Pirc:2013}.

The off case is rather different from the on case, in particular for
small fields. For the smallest applied field of 12\,kV/cm, the EC
response is essentially negligible at all temperatures. In this case
the random initialization with subsequent thermalization in the
applied field has created a FE state for all temperatures below the
ZFH transition temperature and a PE state above
that, and no phase transition occurs under field removal (for the used
$T$-sampling). For the next highest field of 20\,kV/cm there is a sharp
peak with $|\Delta T| \approx 2.6$\,K at $T_\text{i} = 304$\,K, which
is the only temperature for which a phase transition occurs under
field removal for this initial field strength ({\it cf.}
Fig.~\ref{fig:pol}). This case is analogous to the case
shown in Fig.~\ref{fig:delT_T_E20}(a) and already discussed above
(note that for 20\,kV/cm the width of the coexistence region at the
C-T transition is only about 2\,K and thus the initialization is not
very crucial). For fields larger than 20\,kV/cm, the field strength
approaches the critical field, above which
there is no clear difference  between FE and PE
states. Instead, the system is highly polarizable and the EC peak
broadens towards the higher temperature side
for higher fields,  similar to the on case. Thus, for fields above $E_{\text{c}}$ the EC response becomes
fully reversible, in the sense of Eq.~\eqref{eq:EC-cond}, and the
configurational (continuous) part of the EC effect becomes dominant
over the part related to the FO transition entropy.

Towards the low-T side the EC peak exhibits a sharp drop, which is
related to the stabilization of the FE phase for zero field. In the off case, this
occurs when the system cools down during field removal and enters the
coexistence region without undergoing a phase transition to the PE
state. Since the configurational contribution to $\Delta
T$ increases with increasing field strength, this low-$T$ side of the
EC peak is shifted towards higher temperatures with increasing field.
Note that this shift is overestimated in our simulations due to
the overestimation of the unscaled $\Delta T$, as already discussed in
Sec.~\ref{sec:comp-details}.
This effect also leads to the non-monotonous behavior of $\Delta P$ at
$T=304$\,K seen in Fig.~\ref{fig:pol}(b), where for fields above
25\,kV/cm the system remains in the FE state under field removal.

Summarizing the discussion related to Fig.~\ref{fig:ece}, we find a
reversible EC response for fields above the critical field strength
for the FO PE-FE phase transition, whereas a strong difference (both in peak
height and width) is observed between the on and off cases for small
fields. This is in good agreement with experimental observations for
BTO single crystals presented in Ref.~\onlinecite{Moya_et_al_2013}.

\begin{figure}[tb]%
\includegraphics*[width=0.48\textwidth]{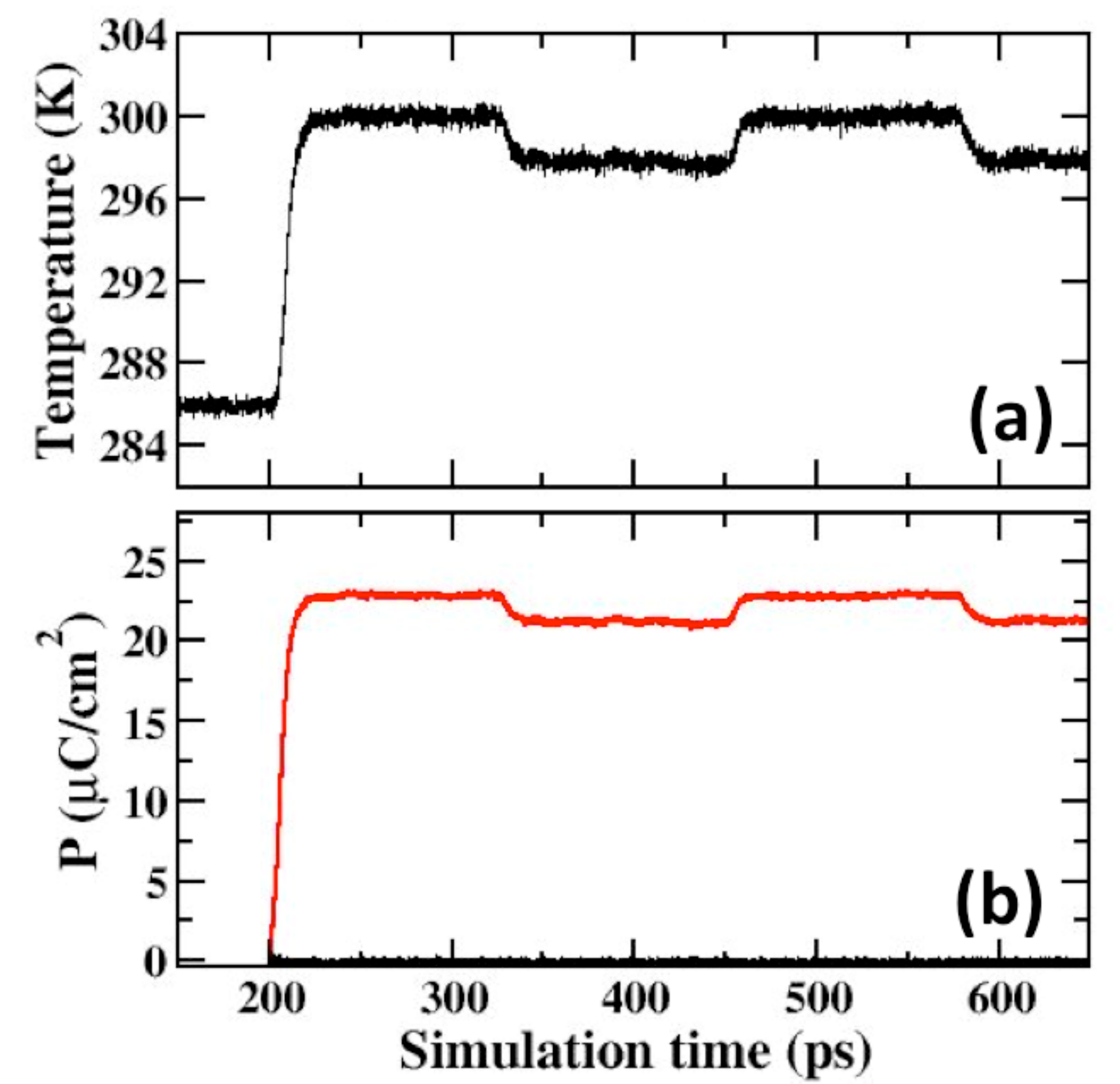}
\caption{Time evolution of the system temperature and polarization
  plotted during periodic cycling of a field of $E_\text{app}$ =
  12\,kV/cm for $T_i$ = 286\,K. The system in initially in the PE-C
  phase. With the application of the field, there is a transition to
  the FE-T phase, leading to a large $\Delta T$ of 2.8\,K (after
  rescaling). On subsequent removal of the field, the system remains
  in the FE-T phase and the EC response is much smaller.}
\label{fig:field-cycling}
\end{figure}

Finally, we examine the (ir-)reversibility of the EC response under field
cycling near the maximum EC response close to the PE-FE (C-T)
transition. Fig.~\ref{fig:field-cycling} shows the time evolution of
the system temperature and polarization as an external field of
$E_{\text{app}} = 12$\,kV/cm is applied and removed periodically during
the course of the MD simulation starting from an initial temperature
of $T_i = 286$\,K. Here, the system is initially in the PE-C phase,
which transforms to the FE-T phase during the first application (ramp
on) of the field. This results in a large EC temperature change,
originating from the transitional contribution due to the FO phase
transition (plus a small configurational part). However, on further
field cycling the system always stays within the coexistence region,
both with and without applied field, and thus it remains in the FE
phase. Consequently, one observes only a much smaller EC response
related to the configurational contribution. 
Thus, an irreversible heating of the system occurs during the first
field field pulse, whereas a reversible but much smaller response
appears during subsequent cycles.
This is analogous to the magnetocaloric response observed at FO
magneto-structural transitions, e.g., in Heusler
alloys.\cite{Titov} There, a large response is found only during the
first field pulse, followed by a small but reversible response in
subsequent cycles.

\section{Conclusions and outlook}

We have used {\it{ab initio}}-based MD simulations for the
prototypical ferroelectric material BTO to demonstrate several
examples of irreversibility and the impact of thermal hysteresis on the
EC response close to FO phase transitions.
We found a large EC response, even for small applied electric fields,
for cases where the system undergoes a FO phase transition. However, this
can depend strongly on the thermal history of the sample and whether
the EC effect is measured under field application or removal.

Although at the C-T transition the coexistence region is narrow, a
large difference between the EC responses observed under field
application and removal, respectively, can occur at small fields. In
particular, a large irreversible heating can arise during the first
field application. This shows that it is important to measure the EC
response for a well prepared initial state either under field removal or
under field cycling in order to obtain reliable results for the
reversible response.

At the T-O transition, the coexistence region is much broader,
resulting in a nearly vanishing response in the FH case, since a small
field is not sufficient to trigger the FO phase transition from the O
to the T phase. Thus, a reversible cycling across the T-O transition
requires significantly larger field strengths.

Generally, for small fields the main contributions to the EC response
results from the transition entropy when crossing the FO phase
transition, whereas the configurational part is small but increases
with increasing field strength.
For the PE-FE (C-T) transition, the configurational part becomes
dominant above the critical field strength $E_\text{c}$, and at
temperatures above the ZFH transition temperature. Therefore,
hysteretic effects can be avoided by using sufficiently large applied
fields, or by cycling the system above the ZFH transition temperature.
In contrast, the configurational contribution to the EC effect is
small within the FE phases near the T-O transition. We also note that, while a reversible
cycling across the T-O transition is in principle possible for very
large fields, a partial cancellation between transitional and
configurational parts can occur, since these two contributions can
have opposite signs.\cite{Marathe_et_al:2017}

We note that the EC effect at the T-O transition shows several
analogies to the well known giant magnetocaloric effect found at
magneto-structural phase transitions, e.g. in Heusler alloys.
\cite{Gutfleisch}  There, a giant response is typically found for the
first field pulse only, and the system stays within the coexistence
range and shows a much smaller but reversible response under further
field cycling. Furthermore, the structural (transitional) and magnetic
(configurational) contributions to the magnetocaloric effect are
opposite in sign,\cite{Gottschall} similar to what can also occur a
the T-O transition in BTO, as stated above.
For the case of the magnetic materials, it has been shown that it is
possible to bypass the hysteresis through an additional stimulus, such
as, e.g., hydrostatic pressure.\cite{Liu2} The same could also be
possible for EC materials exhibiting strong hysteretic effects, due to
the strong coupling of the FE degrees of freedom to pressure or
strain.

Since our model is based on a defect-free homogeneous system, it
allows to clearly isolate intrinsic and different extrinsic factors (such as, e.g., 
domains, defects, inhomogeneities, \dots) determining
hysteretic effects. In the present work we have focused on intrinsic
properties, while the influence of domains, defects, and
inhomogeneities can be incorporated in future studies.  Our simulations
therefore provide a good reference for comparison and analysis of future
experimental investigations of hysteretic effect in EC materials.

\section*{Acknowledgement}
This work was supported by the Swiss National Science Foundation and 
the German Science Foundation under the priority program SPP 1599 
(``Ferroic Cooling").

\bibliography{references}
\end{document}